\documentclass[prb,showpacs,twocolumn,aps,superscriptaddress]{revtex4}
\usepackage{graphicx}
\usepackage{color}
\usepackage{dcolumn}
\usepackage{amsmath}
\usepackage{amssymb}

\newcommand{\lco}{$\rm La_2CuO_4$}

\newcommand{\lsco}{$\rm La_{2-{\it x}}Sr_{\it x}CuO_4$}

\newcommand{\lbco}{$\rm La_{2-{\it x}}Ba_{\it x}CuO_4$}
\newcommand{\lbcon}{$\rm La_{1.905}Ba_{0.095}CuO_4$}
\newcommand{\lbcoe}{$\rm La_{1.875}Ba_{0.125}CuO_4$}
\newcommand{\lbcof}{$\rm La_{1.845}Ba_{0.155}CuO_4$}

\newcommand{\lbsco} {$\rm La_{1.875}Ba_{0.125- \it y}Sr_{\it y}CuO_{4}$}

\newcommand{\lnsco} {$\rm La_{1.6- \it x}Nd_{0.4}Sr_{\it x}CuO_{4}$}
\newcommand{\lnscoe} {$\rm La_{1.475}Nd_{0.4}Sr_{0.125}CuO_{4}$}
\newcommand{\lnscoxy}{$\rm La_{2- \textit{x}- \textit{y}}Nd_\textit{y}Sr_\textit{x}CuO_4$}

\newcommand{\lesco}{$\rm La_{1.8- \it x}Eu_{0.2}Sr_{\it x}CuO_4$}

\newcommand{\nsno}{$\rm Nd_{1.67}Sr_{0.33}NiO_{4}$}
\newcommand{\lsno}{$\rm La_{1.67}Sr_{0.33}NiO_{4}$}

\newcommand{\sus}{susceptibility}
\newcommand{\chic}{$\chi_{c}$}
\newcommand{\chiab}{$\chi_{ab}$}

\newcommand{\schi}{$\chi$}

\newcommand{\pla}{$\rm CuO_2$}
\newcommand{\oct}{$\rm CuO_6$}
\newcommand{\tlt}{$T_{\rm LT}$}
\newcommand{\tht}{$T_{\rm HT}$}
\newcommand{\tco}{$T_{\rm CO}$}
\newcommand{\tso}{$T_{\rm SO}$}

\newcommand{\ooo}{(1,0,0)}
\newcommand{\too}{(2,0,0)}
\newcommand{\tos}{(2,0,6)}
\newcommand{\co}{(2+2$\delta$,0,5.5)}
\newcommand{\so}{(0.5+$\delta$,0.5,0)}

\include{version}

\begin{document}

\title{Stripe order in superconducting $\bf La_{\it 2-x}Ba_{\it x}CuO_{\it 4}$ for $\bf 0.095 \leq x \leq 0.155$}
\author{M. H\"ucker}
\affiliation{Condensed Matter Physics \&\ Materials Science Department, Brookhaven National Laboratory, Upton, New York 11973, USA}
\author{M. v. Zimmermann}
\affiliation{Hamburger Synchrotronstrahlungslabor at Deutsches Elektronen-Synchrotron DESY, 22603 Hamburg, Germany}
\author{G. D. Gu}
\author{Z. J. Xu}
\author{J. S. Wen}
\author{Guangyong Xu}
\affiliation{Condensed Matter Physics \&\ Materials Science Department, Brookhaven National Laboratory, Upton, New York 11973, USA}
\author{H. J. Kang}
\altaffiliation{Present address: Dept.\ of Physics \&\ Astronomy, Clemson University, Clemson, SC 29634-0978, USA}
\affiliation{NIST Center for Neutron Research, National Institute of Standards and Technology, Gaithersburg, Maryland 20899, USA}
\author{A. Zheludev}
\altaffiliation{Present address: Laboratorium f\"ur Festk\"orperphysik, ETH H\"onggerberg 8093 Z\"urich, Switzerland and Laboratory for Neutron Scattering, ETH and Paul Scherrer Institut, Villigen, Switzerland.}
\affiliation{Neutron Scattering Science Division, Oak Ridge National Laboratory, Oak Ridge, Tennessee 37831, USA}
\author{J. M. Tranquada}
\affiliation{Condensed Matter Physics \&\ Materials Science Department, Brookhaven National Laboratory, Upton, New York 11973, USA}

\date{\today}

\begin{abstract}

The correlations between stripe order, superconductivity, and crystal structure in
\lbco\ single crystals have been studied by means of x-ray and neutron diffraction
as well as static magnetization measurements. The derived phase diagram shows that
charge stripe order (CO) coexists with bulk superconductivity in a broad range of
doping around $\rm x=1/8$, although the CO order parameter falls off quickly for
$x\neq 1/8$. Except for $x=0.155$, the onset of CO always coincides with the
transition between the orthorhombic and the tetragonal low temperature structures.
The CO transition evolves from a sharp drop at low $x$ to a more gradual transition
at higher $x$, eventually falling below the structural phase boundary for optimum
doping. With respect to the interlayer CO correlations, we find no qualitative
change of the stripe stacking order as a function of doping, and in-plane and
out-of-plane correlations disappear simultaneously at the transition. Similarly to
the CO, the spin stripe order (SO) is also most pronounced at $x=1/8$. Truly static
SO sets in below the CO and coincides with the first appearance of in-plane
superconducting correlations at temperatures significantly above the bulk transition
to superconductivity (SC). Indications that bulk SC causes a reduction of the spin
or charge stripe order could not be identified. We argue that CO is the dominant
order that is compatible with SC pairing but competes with SC phase coherence.
Comparing our results with data from the literature, we find good agreement if all
results are plotted as a function of $x'$ instead of the nominal $x$, where $x'$
represents an estimate of the actual Ba content, extracted from the doping
dependence of the structural transition between the orthorhombic phase and the
tetragonal high-temperature phase.
\end{abstract}


\maketitle

\section{Introduction}
\label{intro}
The prototypical high-temperature superconductor\cite{Bednorz86} \lbco\ is
particularly well known for its unique doping dependence of the bulk superconducting
(SC) phase.~\cite{Moodenbaugh88a} While its sister compound \lsco , like most other
high temperature superconductors, displays a dome shaped SC phase boundary
$T_c(x)$,~\cite{Takagi89,Broun08a} in the Ba-based compound $T_c(x)$ shows a deep
depression centered at $x=1/8$.~\cite{Moodenbaugh88a,Yamada92}  It was discovered
early on that the so-called 1/8-anomaly is accompanied by a structural transition
from low-temperature orthorhombic (LTO) to low-temperature tetragonal (LTT)
symmetry,\cite{Axe89,Billinge93a} not observed in pure \lsco , and that bulk SC is
replaced by some kind of antiferromagnetic (AF)
order.~\cite{Kumagai91a,Luke91,Sera89,Arai03a} The complex nature of the magnetic
phase was first identified by neutron and x-ray diffraction experiments for an
analogous phase in \lnsco
,~\cite{Tranquada95a,Niemoeller99a,Zimmermann98,Ichikawa00a} and later on confirmed
in \lbsco ,~\cite{Fujita02b,Kimura03b} \lbcoe ,~\cite{Fujita04a,Abbamonte05a,Kim08a}
and \lesco .~\cite{Huecker07b,Fink09a} Undoped ($x=0$), all of these compounds are
quasi two-dimensional commensurate spin $S=1/2$ Heisenberg
antiferromagnets.~\cite{Thio88,Crawford93,Huecker04b}
\begin{figure}[t]
\center{\includegraphics[width=0.95\columnwidth,angle=0,clip]{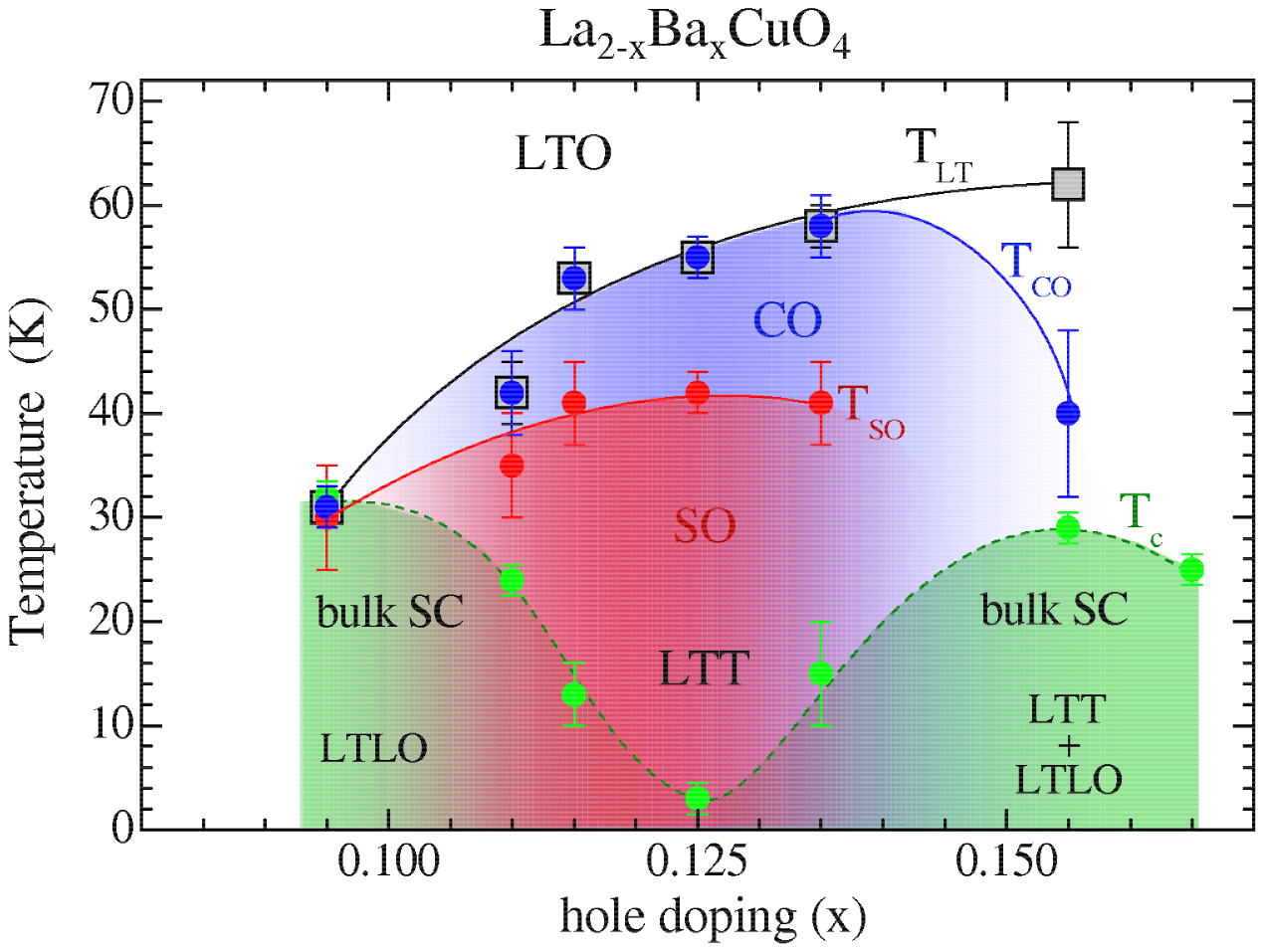}}
\caption[]{(color online) Temperature versus hole-doping phase diagram of \lbco\
single crystals. Onset temperatures: $T_c$ of bulk superconductivity (SC), \tco\ of
charge stripe order (CO), \tso\ of spin stripe order (SO), and \tlt\ of the low
temperature structural phases LTT and LTLO. At base temperature CO, SO, and SC
coexist at least for the crystals with $0.095 \leq x \leq 0.135$. For $x=0.155$ we
identified CO but not SO, and observe a mixed LTT/LTLO phase. In the case of
$x=0.095$ very weak orthorhombic strain persists at low $T$. For $x=0.165$ we have
measured $T_c$ only, before the crystal decomposed. Solid and dashed lines are
guides to the eye. Although \tco , \tso , and \tlt\ for several $x$ were also
determined with XRD and ND, most data points in this figure are from magnetic \sus\
measurements. Here, only \tso\ for $x=0.095$ is from ND and \tco\ and \tlt\ for
$x=0.155$ from XRD.}\label{fig1}
\end{figure}
But doped with sufficient charge carriers, they exhibit incommensurate nuclear and
magnetic superstructure reflections (which we will describe below). Among the
debated interpretations is the so-called stripe model in which the charge carriers
in the \pla\ planes segregate into hole rich stripes, thus forming antiphase
boundaries between intermediate spin stripes with locally AF
correlations.~\cite{Tranquada95a,Zaanen89,Kivelson03a,Vojta09a}
In the LTT phase, which breaks the four-fold rotational symmetry of the \pla\
planes, the electron-lattice coupling is believed to play a central role in the
pinning of stripes,~\cite{Crawford91,Buechner94c,Tranquada96a,Pickett91,Kivelson98}
although recent experiments under pressure revealed that stripes can break the
symmetry even in the absence of long range LTT order.~ \cite{Huecker10a}

%
%
So far, \lnsco\ is the only system with stripe-ordered LTT phase where magnetic {\it
and} charge order have been studied with diffraction on {\it both} sides of
$x=1/8$.~\cite{Ichikawa00a} The results were interpreted as indicating that local
magnetic order (rather than charge stripe order) is responsible for the suppression
of bulk SC, and that charge stripes are compatible with SC as long as the magnetic
correlations remain dynamic. More recent experimental and theoretical results on
\lbcoe\ support the revised view that, in principle, static spin and charge stripes
are compatible with SC pairing, but, due to their orthogonal arrangement in adjacent
planes, they compete with superconducting phase
order.~\cite{Li07c,Tranquada08a,Berg07a,Berg09a}

It is desirable to analyze \lbco\ in a broader range of doping to test the
generality of the observations. 
This system has two advantages over rare-earth-doped \lsco: First, only one element
is substituted for La. Second, the Ba$^{2+}$ ions are non-magnetic, in contrast to,
{\it e.g.}, the Nd$^{3+}$ ions whose large magnetic moments interact with the spins of the
Cu$^{2+}$ ions in the \pla\ planes.~\cite{Wakimoto03a,Nachumi98b} Recent progress in
the synthesis of \lbco\ single crystals with $x \leq 1/8$ has triggered numerous
studies on the stripe order in the underdoped
regime.~\cite{Abe99a,Adachi01a,Fujita04a,Adachi05a,Abbamonte05a,Valla06a,Xu07a,Zhao07a,
Kim08a,Dunsiger08a,Dunsiger08b,Kim09a,Adachi09a}

Despite previous work, however, the doping dependence of many properties requires
further clarification, such as the absolute intensities of CO and SO satellite
reflections, the stripe correlations between the planes, the melting of the stripe
order, and the compatibility with the generic stripe phase diagram. Furthermore,
there is a great lack of information for $x > 1/8$ because crystal growth becomes
progressively more challenging with increasing $x$.

These are the issues addressed in the present study on \lbco\ single crystals with
$0.095 \leq x \leq 0.155$. We have characterized the charge stripe order with
high-energy single-crystal x-ray diffraction (XRD), by probing the associated lattice
modulation.~\cite{Zimmermann98, Niemoeller99a,Kimura03b} That a modulation of the
electron density truly exists, has previously been demonstrated in
Ref.~\onlinecite{Abbamonte05a} for \lbcoe\ by means of resonant soft x-ray
scattering. We have investigated the spin stripe order both in the traditional way, with
neutron diffraction (ND), as well as in a less conventional way by tracing a
recently identified weak ferromagnetic contribution to the normal state magnetic
\sus .~\cite{Huecker08a} The various structural phases have been studied mostly with XRD,
and to some extent with ND, and the SC phase with shielding and Meissner fraction
measurements. As a result, we obtain the temperature versus Ba-concentration phase diagram
displayed in Fig.~\ref{fig1}. One of the key features is that charge stripe order exists over
the entire range of $x$ that we have studied, including the two bulk SC crystals with the lowest and
highest $x$ and maximum $T_c$ on the order of 30~K. According to our quantitative
analysis, the stripe order for these end compositions is already extremely weak,
while it is most pronounced at $x=1/8$. In the underdoped regime the CO always
disappears at the low temperature structural transition, and for three crystals we
can show that it melts isotropically. On the other hand, the onset of bulk SC left
no noticeable mark in our CO and SO data.

The rest of the paper is organized as follows: In Sec.~\ref{exp} we describe the
experimental methods and the choice of reciprocal lattice used to index the
reflections. In Sec.~\ref{results} we present four subsections dedicated to our
results on crystal structure, charge stripe order, spin stripe order, and
superconductivity. In Sec.~\ref{discussion} we summarize the doping dependence of
the various properties as a function of the nominal and an estimated actual Ba
content, compare our results with the literature, and in Sec.~\ref{summary} finish
with a short conclusion.

\section{Experimental}
\label{exp}

\begin{figure}[b]
   \center{\includegraphics[width=0.98\columnwidth,angle=0,clip]{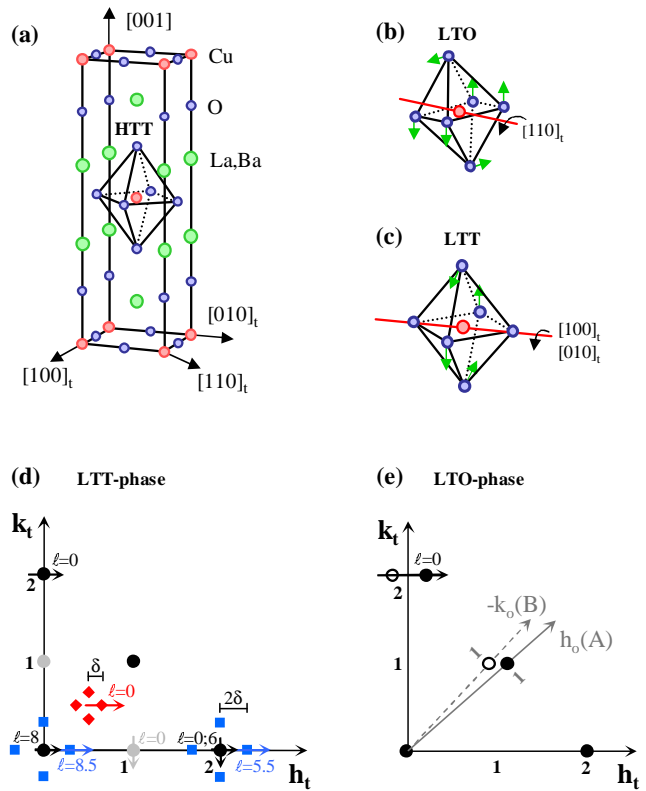}}
   \caption{(color online) Crystal structure and reciprocal lattice of \lbco . (a) Unit cell in
   the HTT phase ($I_4/mmm$). Tilt directions of the CuO$_6$ octahedra in (b)
   the LTO phase ($Bmab$) and (c) the LTT phase ($P4_2/ncm$). Note that in the
   LTT phase the tilt direction alternates between [100]$_t$ and [010]$_t$
   in adjacent layers. The same is true for the stripe direction.
   Reciprocal lattice in terms of the HTT unit cell for (d) the LTT phase and (e)
   the LTO phase, projected along $\ell$ onto the $(h,k)$-plane. Only reflections
   relevant to this work are shown. Fundamental Bragg
   reflections are indicated by black bullets and circles,
   CO reflections by blue squares, SO reflections by red diamonds, and superstructure
   reflections for $\ell = 0$ that are only allowed in the LTT and LTLO phases
   by gray bullets. In (e) we also indicate the reciprocal lattice
   of the orthorhombic phase with its two twin domains A (closed symbols) and B (open symbols).
   The trajectories of typical scans are indicated by arrows, along with the value of
   $\ell$. The HTT phase compares to (d) with only the fundamental Bragg reflections
   present.}\label{fig2}
\end{figure}

A series of six \lbco\ single crystals with $0.095 \leq x \leq 0.155$ has been grown
at Brookhaven with the traveling-solvent floating-zone method.
Previously reported results on some of the compositions, in particular on the
$x=1/8$ crystal, have demonstrated a very high sample quality.~\cite{Tranquada04a,
Valla06a,Homes06a,Li07c,Kim08a,Kim08b,Huecker08a,Tranquada08a,He09a,Huecker10a}
Since the composition of single crystals can deviate from their nominal
stoichiometry, it has been vital to measure structure, stripe order, and SC on
pieces of the same crystal.
In Fig.~\ref{fig2}(a) we show the unit cell of the high-temperature tetragonal (HTT)
phase, with space group $I_4/mmm$. Although the supercells of the low temperature
phases LTO (space group $Bmab$) and LTT (space group $P4_2/ncm$) have a $\sqrt{2}
\times \sqrt{2}$ larger basal plane rotated by 45$^\circ$ degrees, we nevertheless
specify the scattering vectors ${\bf Q}=(h,k,\ell)$ in all phases in units of
$(2\pi/a_t, 2\pi/a_t, 2\pi/c)$ of the HTT cell with lattice parameters $a_t\simeq
3.78$~$\rm \AA $ and $c\simeq 13.2$~$\rm \AA $.~\cite{Maeno91a} In order to express
the orthorhombic strain $s$ in the LTO phase, we will refer to the lattice constants
$a_o$ and $b_o$ of the LTO supercell, which are larger than $a_t$ by a factor of
$\sim\sqrt 2$.

The XRD experiments were performed with the triple-axis diffractometer at wiggler
beamline BW5 at DESY.~\cite{Bouchard98} To create optimum conditions for studying
the bulk properties in transmission geometry, most samples were disk shaped with a
diameter ($\sim 5$~mm) significantly larger than the beam size of $1\times
1$~mm$^2$, and a thickness ($\sim 1$~mm) close to the penetration depth of the
100~keV photons ($\lambda = 0.124$~$ \rm \AA $). Count rates are normalized to a
storage ring current of 100~mA. To evaluate the $x$-dependence of a superstructure
reflection relative to $x=0.125$, we have normalized its intensity with an
integrated intensity ratio $I(0.125)/I(x)$ of a nearby fundamental Bragg reflection.
For example, to normalize the \ooo\ and \co\ reflections, we have applied the
factors $I_{(200)}(0.125)/I_{(200)}(x)$ and $I_{(206)}(0.125)/I_{(206)}(x)$ of the
\too\ and \tos\ Bragg reflections, respectively.

The ND data for $x=0.115$, 0.125 and 0.135 were collected with the triple-axis
spectrometer SPINS located at the NIST Center for Neutron Research using beam
collimations of $55'$-$80'$-S-$80'$-open (S = sample) with fixed final energy $E_f =
5$~meV. The $x=0.095$ crystal was studied at triple-axis spectrometer HB-1 at the
High Flux Isotope Reactor, Oak Ridge National Laboratory, using beam collimations of
$48'$-$48'$-S-$40'$-$136'$ with $E_f = 14.7$~meV. The cylindrical crystals, with a
typical weight between 5~g and 10~g, were mounted with their $(h, k, 0)$-zone
parallel to the scattering plane. Doping dependencies of intensities were obtained
by normalizing the data with the irradiated sample volume.

The static magnetic \sus\ ($\chi = M/H$) measurements, used to study the spin stripe
phase and the SC phase, were performed with a superconducting quantum interference
device (SQUID) magnetometer for $H\parallel c$ and $H\parallel ab$. For these
experiments crystal pieces, with a typical weight of 0.5~g, were used.
\section{Results}
\label{results}
\subsection{Crystal Structure}
\label{structure}
Since the discovery of superconductivity in \lbco\ in the late
eighties,\cite{Bednorz86} the crystal structure, displayed in Fig.~\ref{fig2}, has
been studied intensively.~\cite{Axe89} So far most diffraction results were obtained
on polycrystals,~\cite{Axe89,Maeno91a,Billinge93a} and only recently have
single-crystal data been reported.~\cite{Fujita04a,Wakimoto06a,Zhao07a}
\begin{figure}[t]
\center{\includegraphics[width=0.82\columnwidth,angle=0,clip]{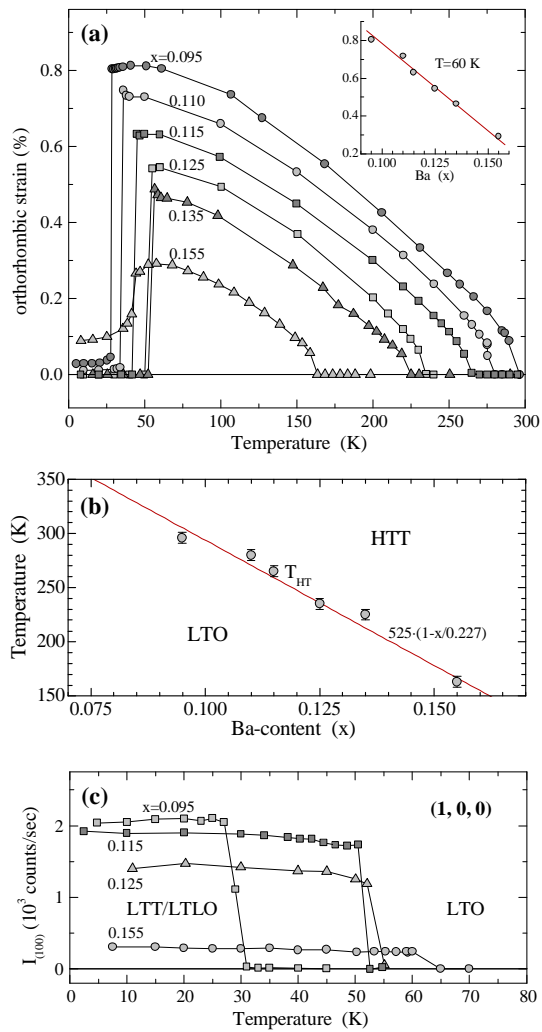}}
\caption[]{(color online) Structural properties of \lbco\ from XRD. (a) Orthorhombic
strain $s$ versus temperature as a function of Ba-doping. $s$ was determined from
transverse scans through the (2,0,0)/(0,2,0) Bragg reflections which are
simultaneously present due to twin domains; see Fig.~\ref{fig2}(e). For $x=0.155$
the LTO$\leftrightarrow$LTLO transition of the majority phase is shown, although a
significant volume fraction of 20\% turns LTT at \tlt ; see Fig.~\ref{fig12}. Inset:
$s$ versus $x$ at 60~K. The solid line is a guide to the eye. (b) Temperature of the
HTT$\leftrightarrow$LTO transition versus doping. The solid line describes $T_{\rm
HT}(x)$ using $T_{\rm HT}(0) \times (1-x/x_c)$. $T_{\rm HT}(0)$ and $x_c$ were
chosen such that at $x=0$ the line intercepts at 525~K for \lco\ and goes through
235~K at $x=0.125$, which is the most accurately known $T_{\rm HT}$ for Ba-doped
compounds. (c) Integrated intensity from $k$-scans through the (1,0,0)
superstructure peak. For (a) and (c) error bars are within symbol size.}
\label{fig3}
\end{figure}
In the doping range considered here, \lbco\ undergoes two structural transitions
with decreasing temperature: a second-order transition from HTT to LTO, and a
first-order transition from LTO to another low temperature phase which can either be
LTT or the low-temperature less-orthorhombic (LTLO) phase (space group $Pccn$) which
is a possible intermediate phase between LTO and LTT.~\cite{Crawford91} While the
HTT phase is characterized by untilted \oct\ octahedra forming flat \pla\ planes,
all low-temperature phases can be described by different patterns of tilted \oct\
octahedra;  see Fig.~\ref{fig2}(a-c).~\cite{Buechner91,Crawford91,Maeno91a} In the
LTO phase,the octahedra tilt by an angle $\Phi$ about the tetragonal [1,1,0]$_t$
axis which is diagonal to the \pla\ square lattice and defines the orthorhombic
[1,0,0]$_o$ axis [Fig.~\ref{fig2}(b)]. In the LTT phase, the tilt axis runs parallel
to the square lattice, but its direction alternates between [1,0,0]$_t$ and
[0,1,0]$_t$ in adjacent planes.~\cite{Axe89,Maeno91a,Tranquada95a} In the LTLO
phase, the tilt axis points along an intermediate in-plane
direction.~\cite{Crawford91}

The structural properties in this section were obtained with XRD, while data from ND
are presented in Sec.~\ref{neutron}. In Fig.~\ref{fig3}(a) we show, for all $x$, the
temperature dependence of the orthorhombic strain $s=2(b_o-a_o)/(a_o+b_o)$, from
which we have extracted the HTT$\leftrightarrow$LTO transition temperature, \tht, as
a function of doping. The maximum strain $s$ the lattice reaches at low temperatures
is directly, although nonlinearly, related to \tht .~\cite{Buechner94c} Both
quantities show a monotonic decrease with increasing $x$, as shown in the inset to
Fig.~\ref{fig3}(a) and in Fig.~\ref{fig3}(b). In particular, we observe that \tht\
decreases at a rate of $dT_{\rm HT}/dx$ of $\sim$23.1 K/0.01 Ba [solid line in
Fig.~\ref{fig3}(b)], which is very similar to published polycrystal
data.~\cite{Axe89,Adachi01a,Zhao07a} The difference between a crystal's \tht\ value
and this line can be used to estimate the deviation of its actual Ba concentration
$x'$ from the nominal $x$. Overall the data in Fig.~\ref{fig3}(b) show that $x$ is a
fairly good representation of $x'$. Nevertheless, in the discussion in
Sec.~\ref{discussion} we will show that small discrepancies between our results and
data in the literature can be reconciled in terms of $x'$.

The second transition, at \tlt, from LTO to either LTT or LTLO, causes a sudden drop
of the orthorhombic strain at low temperatures, as one can see in
Fig.~\ref{fig3}(a). In particular, for $x=0.115$, 0.125 and 0.135 we observe
discontinuous LTO$\leftrightarrow$LTT transitions. The crystals with $x=0.11$ and
0.095 show discontinuous LTO$\leftrightarrow$LTLO transitions with very weak strain
remaining below \tlt; the strain continues to decrease at low temperatures and, for
$x=0.11$, eventually becomes zero. The crystal with $x=0.155$ shows a discontinuous
transition that results in a mixed LTLO/LTT phase, as is discussed in more detail in
Sec.~\ref{special}. (That crystal also consisted of several domains, but we were
able to isolate the diffracted signal from a single domain region.)

To examine the low-temperature transition in more detail, we have followed the
temperature dependence of the (1,0,0) superstructure reflection, which is allowed in
the LTT and LTLO phases, but not in the LTO phase. In Fig.~\ref{fig3}(c) we show
integrated intensities $I_{(100)}$ normalized with the (2,0,0) Bragg reflection as
previously explained. As $x$ increases, one can see that $I_{(100)}$ drops while
$T_{\rm LT}$ grows. This behavior indicates that local structural parameters are
involved in the mechanism that drives the transition, as will be discussed further
in Sec.~\ref{discussion}.

At this point we mention that the low temperature transition is also visible in the
static magnetic \sus\ for dopings $x \leq 0.135$, and we find good agreement with
the diffraction data for \tlt ; see Fig.~\ref{fig8} and Ref.~\onlinecite{displex}.

\subsection{Charge Stripe Order}
\label{charge}
\begin{figure}[t]
\center{\includegraphics[width=0.6\columnwidth,angle=0,clip]{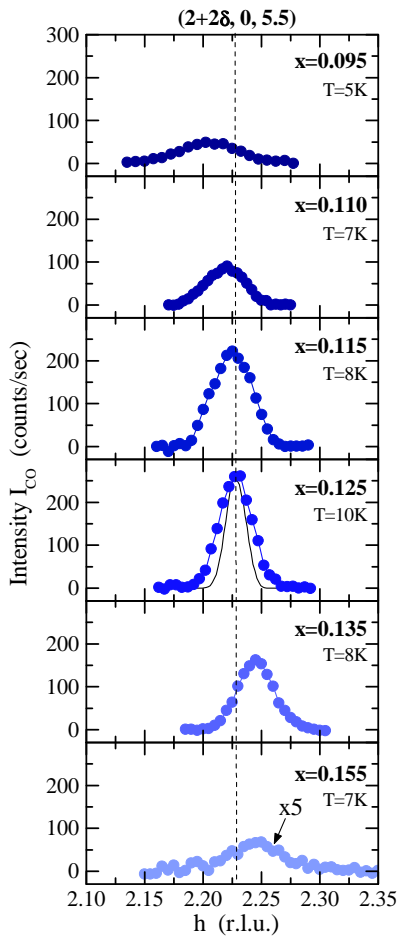}}
\caption[]{(color online) In-plane CO correlations. $h$-scans through the CO-peak at
\co\ for different dopings after subtraction of a linear background. The intensities
are normalized to the integrated intensity of the \tos\ Bragg reflection, as
explained in Sec.~\ref{exp}, and are directly comparable. Error bars are within
symbol size. The data for $x=0.155$ has been multiplied by a factor of five. The
dashed line marks the CO-peak position for $x=0.125$ and emphasize its shift with
doping. All scans were collected at \co\ in the same scattering geometry. The
corresponding resolution function was measured at the (2,0,6) Bragg reflection and
is indicated by a solid line for $x=0.125$.}\label{fig4}
\end{figure}

The charge stripe order, studied with XRD, leads to weak reflections with ordering
wave vectors ${\bf Q}_{\rm CO}=(2\delta,0,0.5)$ and $(0,2\delta,0.5)$, where
$\delta$ increases with hole concentration; see
Fig.~\ref{fig2}(d).~\cite{Tranquada97a,Yamada98a}. In Fig.~\ref{fig4} we show
$h$-scans through the (2+2$\delta$,0,5.5) CO-peak for different $x$ at base
temperature. To accurately determine the position and intensity, we used the (2,0,6)
Bragg reflection as a reference. All these scans were performed with identical
scattering geometry, for which we kept the [0,1,0]$_t$ direction in the scattering
plane. This guaranties the same relative orientation in $k$-space of the CO-peak and
the resolution ellipsoid, which has been determined at the (2,0,6) Bragg peak; in
Fig.~\ref{fig4}, the resolution limited peak shape along $h$ is indicated for
$x=0.125$.

\begin{figure}[t]
\center{\includegraphics[width=0.74\columnwidth,angle=0,clip]{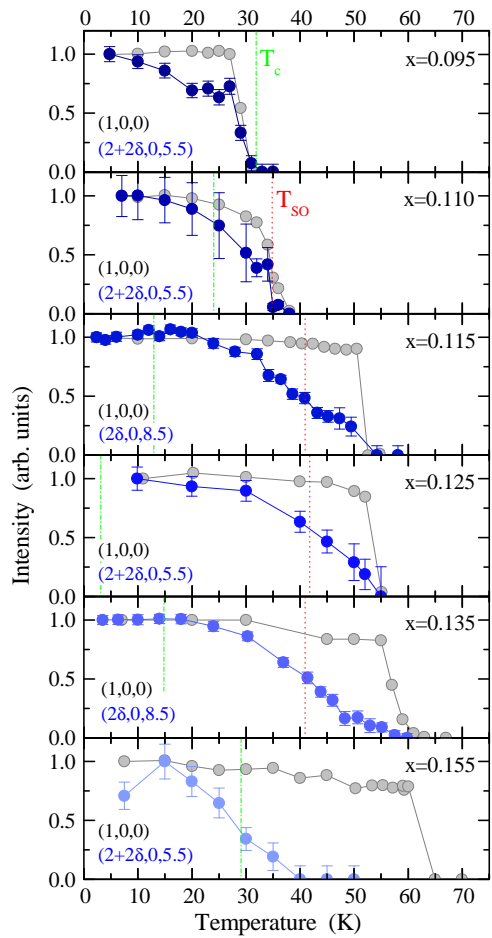}}
\caption[]{(color online) Integrated intensity versus temperature and doping from
$k$-scans through the (1,0,0) peak and $h$-scans through the CO-peak. Note that the
data in this figure was not measured for all dopings in the same scattering geometry
and in the case of the CO-peak not always at the same $(h,k,l)$ position. Therefore,
presented intensities are normalized at low temperature. Error bars for the (1,0,0)
intensity are within symbol size. The green dash-dotted lines indicate the onset of
bulk SC at $T_c$ and the red dotted lines the onset of SO at \tso .}\label{fig5}
\end{figure}

\begin{figure}[t]
\center{\includegraphics[width=0.63\columnwidth,angle=0,clip]{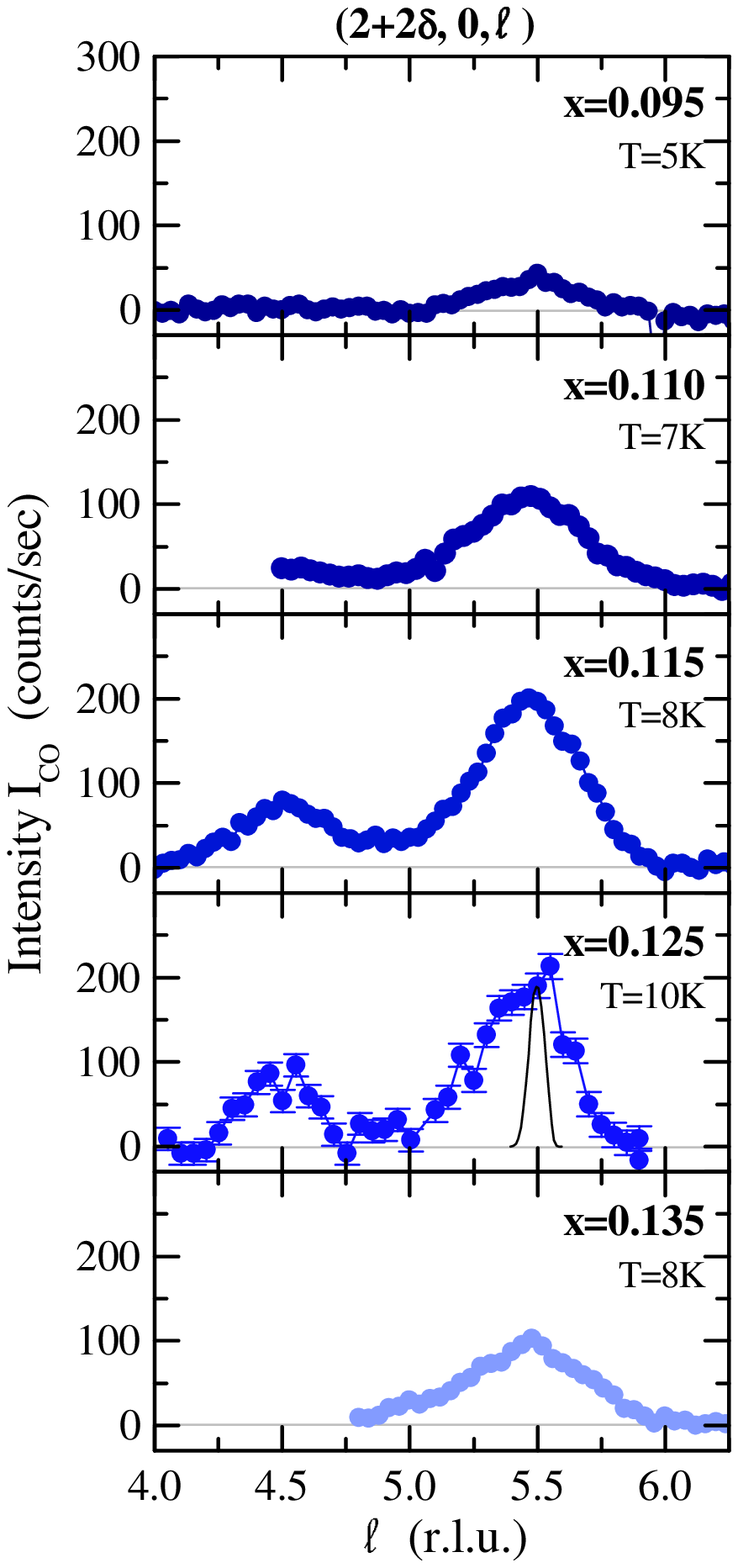}}
\caption[]{(color online) Interlayer CO correlations. $\ell$-scans along ${\bf
Q}=($2+2$\delta$,0,$\ell$) at base temperature for different dopings after
background subtraction. Intensities are normalized to the integrated intensity of
the \tos\ Bragg reflection, as explained in Sec.~\ref{exp}, and are directly
comparable. The data for $x=0.125$ has been taken from Ref.~\onlinecite{Huecker10a}
and was measured in a pressure cell, which explains the low counting statistic. The
data has been corrected for the absorption of the pressure cell. All scans were
collected at \co\ in the same scattering geometry. Error bars are not shown if
within symbol size. The resolution function was measured at the (2,0,6) Bragg
reflection and is indicated by the solid line for $x=0.125$. Because of a
significant doping and temperature dependence of the background, there is no unique
way to subtract it. In most cases the background was either measured along the same
$\bf Q$ at $T \gtrsim T_{\rm LT}$, or along ${\bf Q}=($2+2$\delta$, 0.03, $\ell$) at
base temperature.}\label{fig14}
\end{figure}

As one can see in Fig.~\ref{fig4}, the peak intensity is maximum at $x=1/8$ and
falls off rapidly for $x \neq 1/8$. To our surprise, we still find weak CO-peaks for
dopings as low as $x=0.095$ and as high as $x=0.155$. The incommensurability
$2\delta$ extracted from the peak position shifts monotonically from 0.205 for
$x=0.095$ to 0.245 for $x=0.155$. The empirical relationship\cite{Yamada98a} $\delta
\approx x$ for $x\le 1/8$ would predict $2\delta=0.25$ at $x=0.125$, but the
experimental value clearly stays below, as has been noticed by other groups, as
well.~\cite{Kim09a}

\begin{figure}[t]
\center{\includegraphics[width=0.9\columnwidth,angle=0,clip]{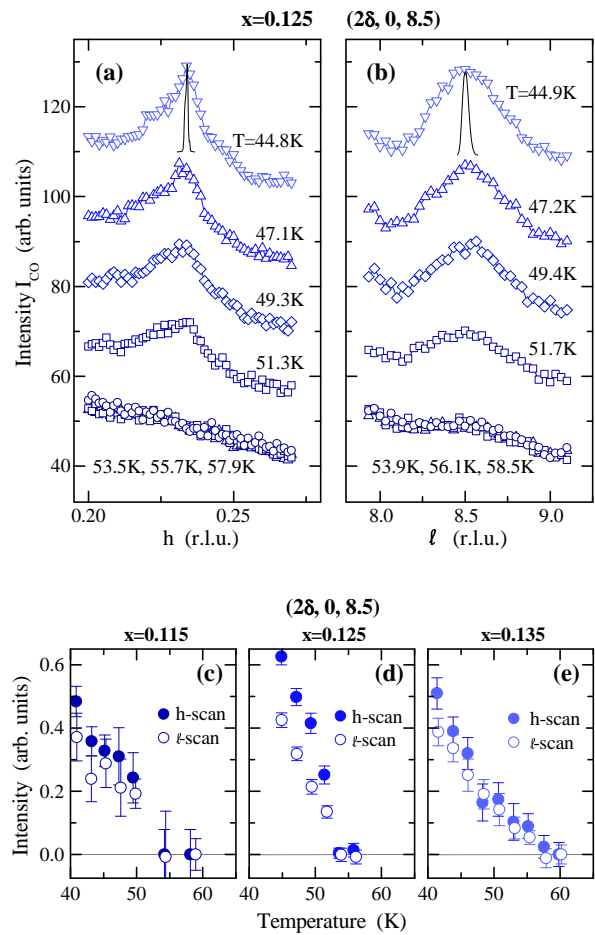}}
\caption[]{(color online) Melting of the charge stripe order. (a) $h$-scans and (b)
$\ell$-scans through the (2$\delta$,0,8.5) CO-peak for $x=0.125$ at different
temperatures. Curves for $T<T_{\rm CO}$ are shifted for clarity. Error bars are
within symbol size. The solid lines indicate the resolution function measured at the
(0,0,8) Bragg reflection. The remaining profile in the $\ell$-scans for $T>T_{\rm
CO}$ originates from diffuse scattering around (0,0,8). (c-e) Integrated intensities
from corresponding $h$ and $\ell$-scans for $x=0.115$, 0.125 and 0.135. The data in
(c-e) was normalized at a base temperature of $\sim 3$~K where additional scans were
performed. }\label{fig15}
\end{figure}

In Fig.~\ref{fig5} we compare the temperature dependence of the CO and the (1,0,0)
peak intensities for the different dopings. This time we show normalized integrated
intensities since not all data sets do correspond to identical reflections,
scattering geometry, or sample thickness. One important finding is that for $x \leq
0.135$ the onset of charge order always coincides with the LTO$\rightarrow$LTT/LTLO
transition, {\it i.e.}, $T_{\rm CO}=T_{\rm LT}$. Only for $x =0.155$ does $T_{\rm
CO}$ drop below \tlt . The temperature dependence of the CO and (1,0,0) peak
intensities evolve differently. Independent of $x$, the (1,0,0) peak shows very
sharp transitions, and is nearly constant below $T_{\rm LT}$. This is indicative of
the transition's first order nature and shows that $\Phi^2 \propto I_{(100)}$ barely
increases at low $T$.
In the case of the CO peak, we see an evolution from a sharp transition at $x=0.095$
to one at $x=0.135$ where CO fades away on warming, until finally at $x=0.155$ we
find $T_{\rm CO}<T_{\rm LT}$. For $x=0.095$ in particular, the data suggest that CO
could persist at higher temperatures if it were not cut off by the structural
transition.

\begin{figure}[t]
\center{\includegraphics[width=0.78\columnwidth,angle=0,clip]{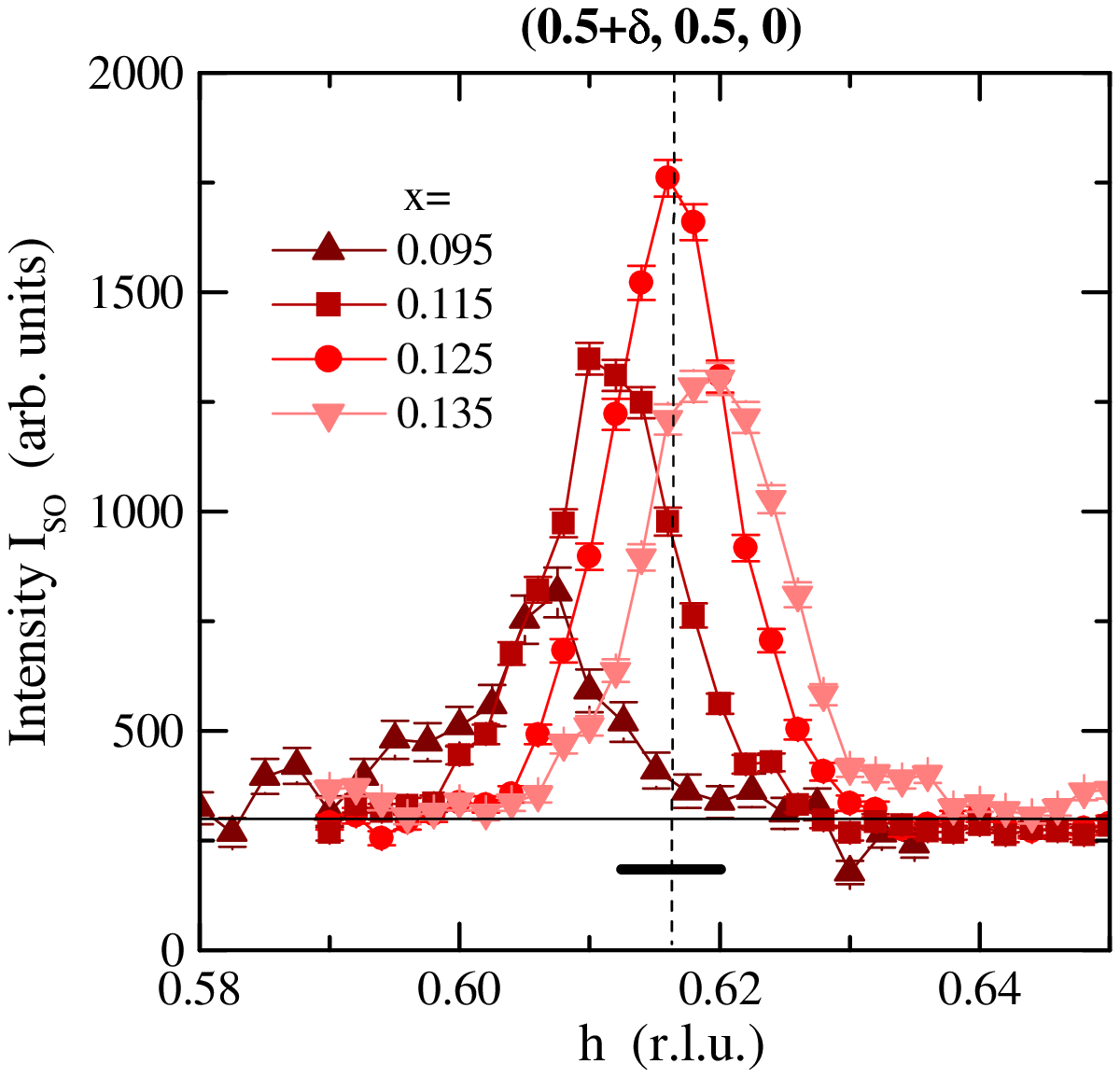}}
\caption[]{(color online) In-plane SO correlations. $h$-scans through the SO-peak at
\so\ for different dopings. The horizontal bar at the bottom indicates the
instrumental resolution full width at half maximum (FWHM). The intensities have been
normalized to the crystal volume in the neutron beam and for $x\geq 0.115$ are
directly comparable; see text and Sec.~\ref{exp}.}\label{fig6}
\end{figure}

\begin{figure}[t]
\center{\includegraphics[width=0.8\columnwidth,angle=0,clip]{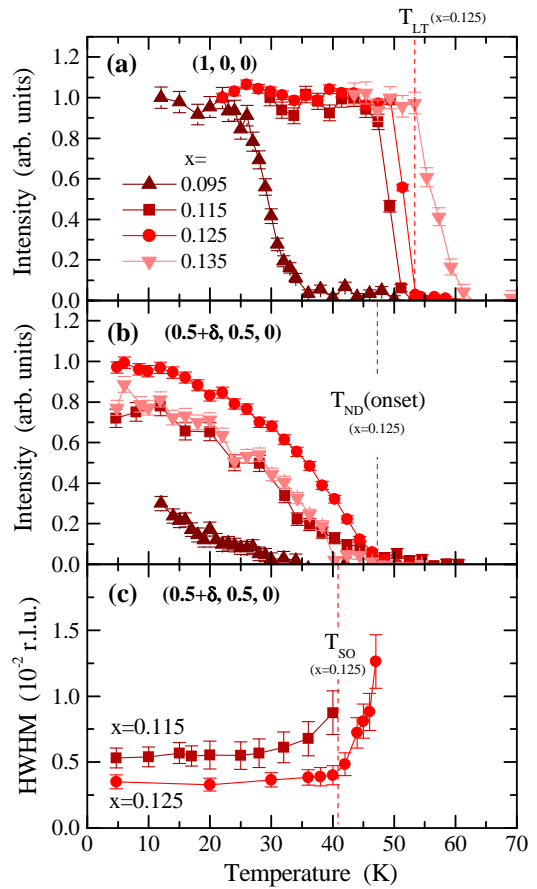}}
\caption[]{(color online) Peak intensity versus temperature and doping of (a) the
\ooo\  superstructure peak normalized at low temperature and (b) the \so\ SO-peak
normalized to the crystal volume in the neutron beam. The data in (b) are directly
comparable only for $x\geq 0.115$; see text and Sec.~\ref{exp}. (c) Resolution
corrected half width at half maximum (HWHM) versus temperature for $x=0.115$ and
$x=0.125$. The dashed lines in (a-c) indicate the onset temperatures for $x=0.125$
of the LTO$\leftrightarrow$LTT transition, the SO-peak intensity, and the peak
broadening. Data for $x=0.125$ previously presented in
Ref.~\onlinecite{Tranquada08a}.}\label{fig7}
\end{figure}

\subsubsection{Charge stripe stacking order}
An important question concerns the doping dependence of the stripe correlations
perpendicular to the \pla\ planes. For $x=1/8$, the stacking arrangement in \lbco\
and \lnsco\ is well known. Stripes run parallel to the Cu-O bonds but in orthogonal
directions in adjacent planes. Thus, only in every other layer do stripes run
parallel, but in addition they are shifted by half the charge period, which results
in a body-centered type of stacking, with a repeat of four planes (two unit
cells).~\cite{Tranquada95a,Zimmermann98,Kimura03b,Kim08a} Therefore, CO-peaks occur
at half integer $\ell$ positions. To test the robustness of this stacking pattern as
a function of hole concentration, we performed the $\ell$-scans shown in
Fig.~\ref{fig14}. Similar to Fig.~\ref{fig4}, the data show absolute intensities
obtained in identical scattering geometry. It is obvious that, in spite of the
dramatic variation of the intensity, all scans show the same modulation in $\ell$.
This clearly demonstrates that the stacking order type is the same in the studied
range $0.095 \leq x \leq 0.135$, and rules out a dramatic change of the correlation
length perpendicular to the planes. Note that for $x=0.155$ the intensity was too
weak to identify the $\ell$ dependence.

\begin{figure*}[t]
\center{\includegraphics[width=2.05\columnwidth,angle=0,clip]{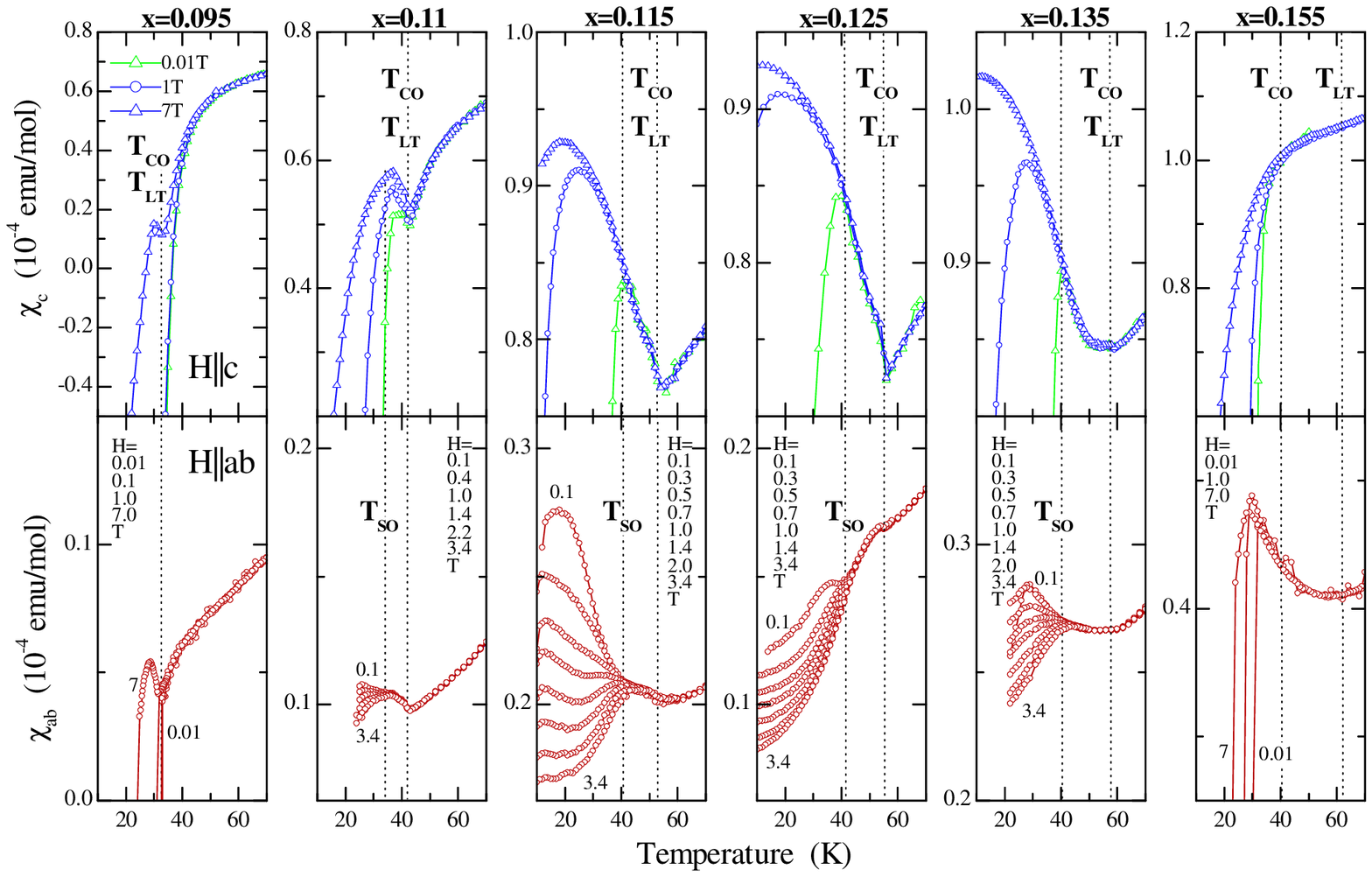}}
\caption[]{(color online) Field cooled static \sus\ $\chi=M/H$ of \lbco\ versus
temperature for different dopings and magnetic fields applied parallel to the
$c$-axis (top) and parallel to the $ab$-plane (bottom). Similar to
Ref.~\onlinecite{Huecker08a}, small deviations between curves for same $x$ but
different fields, due to experimental error ($\pm 0.007 \times 10^{-4}$~emu/mol),
were corrected by shifts in $\chi$, so that curves match for $T > T_{\rm LT}$ where
no field dependence was observed. The core diamagnetism and the Van Vleck \sus\ of
the $\rm Cu^{2+}$ ions have not been subtracted.~\cite{Huecker08a} The dashed lines
mark the onset of SO and CO as well as the LTO$\leftrightarrow$LTT/LTLO transition.
CO leads to anomalies most pronounced for $H\parallel c$. SO is identified by means
of a weak ferromagnetic transition for $H\parallel ab$. One can see that \tso\ also
coincides with the onset of weak diamagnetism from superconducting correlations for
$H\parallel c$. Furthermore, for $x \leq 0.135$ \tco\ coincides with \tlt . For
$x=0.155$ no anomalies are observed and \tco\ and \tlt\ are from XRD. For $x=0.095$
only one transition at $T_{\rm CO}=T_{\rm LT}$ is observed. We have limited the data
for $H\parallel ab$ to fields below the spin-flop transition~\cite{Huecker08a} which
will be the focus of a future publication.}\label{fig8}
\end{figure*}

Another question concerns the way the charge stripe order melts as the temperature
approaches \tco . There is evidence for $x=1/8$, as well as for isostructural
nickelates such as \nsno, that the stacking order melts well before the in-plane
order disappears at \tco .~\cite{Kim08a,Huecker06a} To check if our crystals show
this effect, we performed scans through the (2$\delta$,0,8.5) CO-peak along $h$ and
$\ell$ for $x=0.115$, 0.125 and 0.135; see Fig.~\ref{fig15}. The data in
Fig.~\ref{fig15}(a) and (b) for $x=0.125$ clearly show that the peak in $\ell$
remains well defined until it disappears simultaneously with the peak in $h$. In
Fig.~\ref{fig15}(c-e) we show the integrated intensities of the $h$ and $\ell$ scans
for three dopings, normalized at a base temperature of $\sim 3$~K. One can see that
in all cases the extracted intensities for $h$ and $\ell$ disappear simultaneously.
Thus, we conclude that the stacking order persists up to $T_{\rm CO}$, and that the
CO melts isotropically. We mention that the measurements in Fig.~\ref{fig15} were
performed with the $(h,0,\ell)$ zone parallel to the scattering plane, which gives a
very good resolution, indicated by scans through the (0,0,8) Bragg peak.

\subsection{Spin Stripe Order}
\label{spin}

\subsubsection{Neutron diffraction}
\label{neutron}
The spin stripe order was studied by means of neutron diffraction and static
magnetization measurements. Neutron diffraction allows one to directly probe the
incommensurate spin structure of the SO and thus provides crucial complementary
information to the x-ray diffraction data on the incommensurate CO. The magnetic
ordering wave vectors are ${\bf Q}_{\rm SO}=(0.5-\delta, 0.5, 0)$ and $(0.5,
0.5-\delta, 0)$, {\it i.e.}, they are displaced by $\delta$ from the position of the
magnetic Bragg peak in the AF parent compound \lco , as indicated in
Fig.~\ref{fig2}(d).
%
%
In Fig.~\ref{fig6} we show $h$-scans through the $(0.5+\delta, 0.5, 0)$ SO-peak for
the different dopings. The data for $x \geq 0.115$ was taken at SPINS with identical
configuration and is normalized to the crystal volume in the beam, thus allowing a
direct comparison of the intensities. One can see that the SO-peak is maximum for
$x=1/8$, just as for the CO-peak. The data for $x=0.095$ were taken with the HB-1
spectrometer; they show a SO-peak that is definitely much weaker, although the
available data is insufficient to draw a precise quantitative comparison to the
other dopings. No SO-peak was detected for $x=0.155$, which could be because stripe
order has become very weak. On the other hand, this crystal is a good bulk
superconductor with $T_c=29$~K, so it could be that there is a spin gap below $T_c$
in place of a SO-peak.~\cite{Kofu09a} As one can see in Fig.~\ref{fig6}, the SO-peak
shifts to higher $h$ with increasing $x$, reflecting a similar increase of $\delta$
as for the CO-peak; details will be discussed in Sec.~\ref{discussion}.

\begin{figure}[t]
\center{\includegraphics[width=0.7\columnwidth,angle=0,clip]{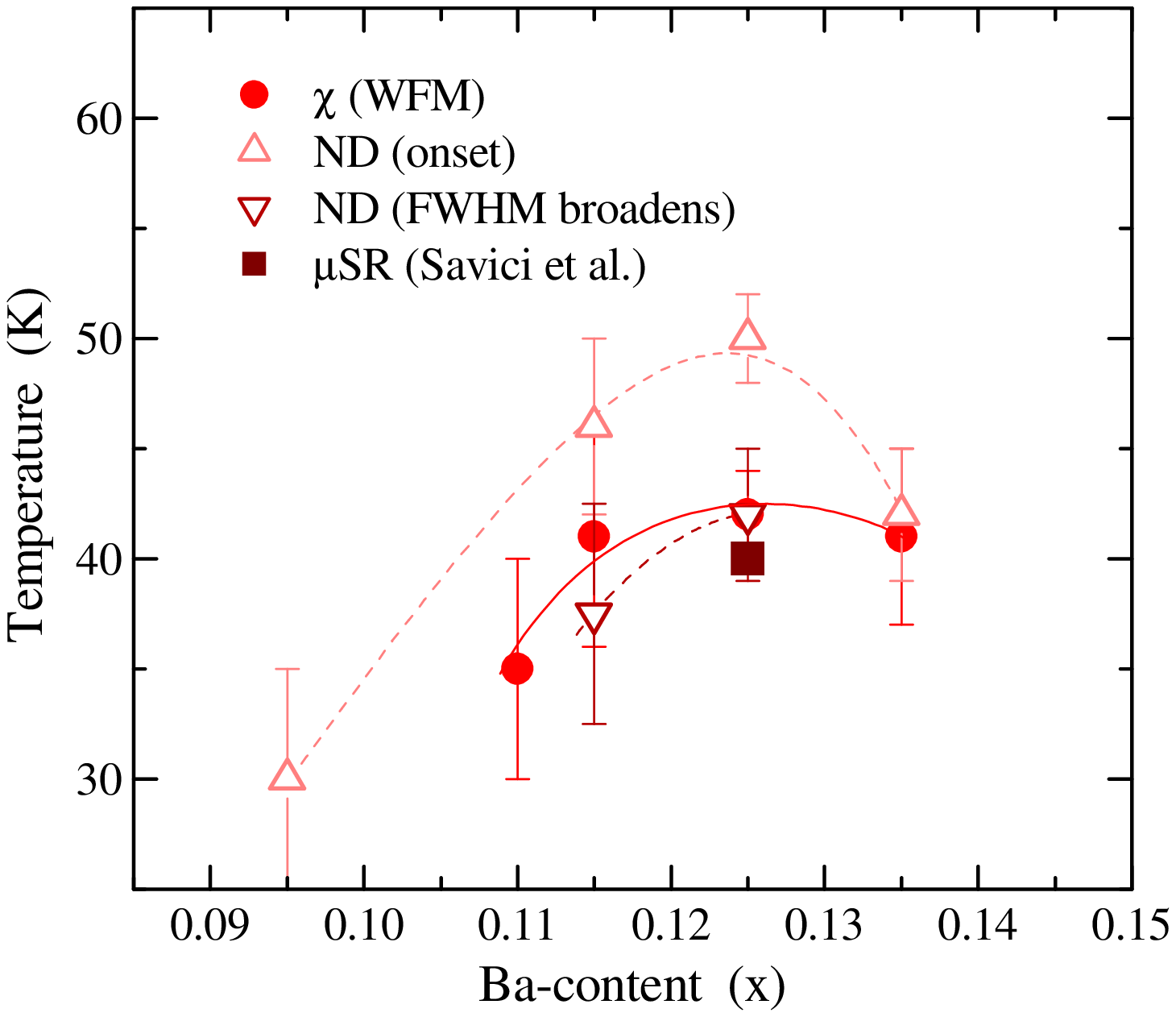}}
\caption[]{(color online) Critical temperatures of the SO transition in \lbco\
single crystals. ($\vartriangle$) Onset of SO-peak intensity and ($\triangledown$)
saturation temperature of SO-peak width as measured with ND, ($\bullet$) onset of
weak ferromagnetism (WFM) for $H\parallel ab$ in static magnetic \sus , which we
associate with $T_{\rm SO}$. ($\blacksquare$) $T_{\rm SO}$ as measured with $\mu$SR;
taken from Ref.~\onlinecite{Savici05a}.} \label{fig9}
\end{figure}

Next, in Fig.~\ref{fig7} we compare the temperature dependence of the peak intensity
of the \ooo\ peak and the SO-peak as measured with ND. The first thing to note is the
good agreement of the \ooo\ data in Fig.~\ref{fig7}(a) with corresponding XRD data
in Fig.~\ref{fig5} regarding $T_{\rm LT}(x)$ and the sharpness of the
LTO$\leftrightarrow$LTT/LTLO transitions, considering the experimental errors
resulting from the use of different instruments, and from the fact that the neutron
beam averages over a sample volume that is three orders of magnitude larger than for
XRD. This indicates a high homogeneity of the crystals' stoichiometry and quality.

Figure~\ref{fig7}(b) shows the SO-peak intensity as a function of temperature. One
can see that for $x \geq 0.115$, intensity sets in at a temperature below $T_{\rm
LT}$. The gap to $T_{\rm LT}$ is particularly wide for $x = 0.135$, but also clearly
visible for $x=0.115$ and 0.125. For $x=0.095$, the onset is about 30~K, but the low
statistics of the weak signal prevent a precise correlation with the other
transitions.

It is known from, {\it e.g.}, $\mu$SR measurements~\cite{Savici05a,Tranquada98b}
that truly static SO sets in below the onset temperature seen by neutron
diffraction. The difference is due to the coarser energy resolution of neutron
diffraction, which can sample critical fluctuations at $T > T_{\rm SO}$. In
Ref.~\onlinecite{Tranquada08a} we have argued, for the case of $x=0.125$, that \tso\
coincides with the temperature above which the SO-peak starts to broaden; see
corresponding data for two dopings in Fig.~\ref{fig7}(c). Furthermore, we have shown
in Refs.~\onlinecite{Li07c,Huecker08a} for $x=0.125$ that \tso\ is also marked by a
weak ferromagnetic transition in the static magnetic \sus\ for magnetic fields
$H\parallel ab$, which we discuss next.~\cite{weakferro}

\subsubsection{Static magnetic \sus}
\label{suscept}
In Fig.~\ref{fig8} we compare the static \sus\ \schi\ for dopings $0.095
\leq x \leq 0.155$. A detailed description of \schi\ in \lbcoe\ has been given in
Ref.~\onlinecite{Huecker08a}. The top panels of Fig.~\ref{fig8} are for $H\parallel
c$ and display the suppression of diamagnetic contributions to the normal state
\sus\ from SC, which leads to an \textit{increase} of \chic\ with field. In
contrast, the bottom panels for $H\parallel ab$ display the weak ferromagnetic
behavior, which is visible for $0.11 \leq x \leq 0.135$ and characterized by a
\textit{decrease} of \chiab\ with field for $T<T_{\rm SO}$. For $x=0.095$ and
$x=0.155$ and $H\parallel ab$ the static \sus\ reveals no information on SO, simply
because the onset of bulk SC has shifted to higher temperatures and obscures any
signature of the normal state weak ferromagnetism.

It is remarkable to see in Fig.~\ref{fig8} that for those $x$ displaying weak
ferromagnetism for $H \parallel ab$, $T_{\rm SO}$ coincides with the temperature
$T_c^*$ of first appearance of superconductivity for $H \parallel c$ in the limit of
small magnetic fields; see dashed lines denoted $T_{\rm SO}$. Note that in
Fig.~\ref{fig8} we are looking at an extremely fine scale. For comparison, $1\times
10^{-4}$~emu/mol equals only $0.0023\%$ of the full Meissner response. In
Refs.~\onlinecite{Li07c,Tranquada08a} we have discussed the idea that in \lbcoe\
this weak diamagnetism in the LTT phase emerges from two-dimensional (2D)
superconducting fluctuations below $T_c^*$, rather than three-dimensional (3D) bulk
superconductivity which sets in at a lower temperature $T_c$ and will be discussed
in Sec.~\ref{SC}. Here we have shown that $T_{\rm SO}=T_c^*$ in a broader range of
doping around $x=1/8$.

\subsubsection{Comparison of critical temperatures}

In Fig.~\ref{fig9} we compare the various critical temperatures of the SO phase,
extracted by ND and \schi\ measurements as well as by $\mu$SR in
Ref.~\onlinecite{Savici05a}. There is good agreement for \tso\ from \schi\ and
$\mu$SR as well as from the SO-peak broadening in ND, whereas the onset temperatures
of finite SO-peak intensity are higher. For the phase diagram in Fig.~\ref{fig1} we
decided to show $T_{\rm SO}$ from \schi , since this is the most complete set of
values. Only for $x=0.095$ did we take the onset temperature from ND, knowing that
truly static SO most likely occurs at a lower temperature. For $0.11 \leq x \leq
0.135$, where we are more confident of the determination, one can see from
Fig.~\ref{fig1} that $T_{\rm SO}$ is always lower than \tco .

\begin{figure}[t]
\center{\includegraphics[width=0.8\columnwidth,angle=0,clip]{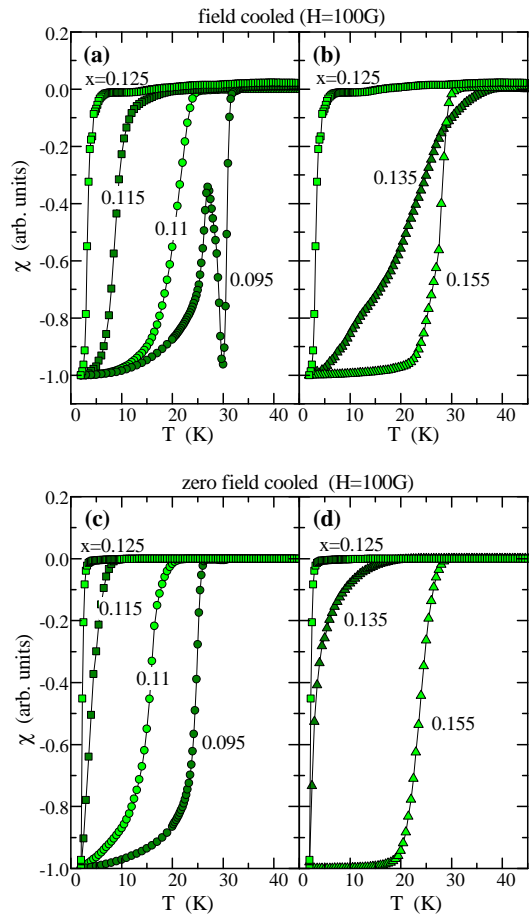}}
\caption[]{(color online) Superconductivity in \lbco . Normalized field cooled (top)
and zero field cooled (bottom) \sus\ versus temperature for a magnetic field of
$H=100$~G applied parallel to the $c$-axis. Data in (a) and (c) are for $x \leq 1/8$
and in (b) and (d) for $x \geq 1/8$.}\label{fig10}
\end{figure}

Returning to Fig.~\ref{fig8}, we mention that \schi\ also shows anomalies at $T_{\rm
LT}$ which are particularly pronounced for $H \parallel c$ and $x\leq 0.125$; see
dashed lines. For these dopings we know that $T_{\rm CO}=T_{\rm LT}$. In contrast,
the anomaly is quite small in the case of $x=0.135$, where \chic\ starts to increase
significantly only for $T<50$~K. This is consistent with the sample's CO in
Fig.~\ref{fig5} which becomes already weak at $T\sim 50$~K before it eventually
disappears at \tco . Since the structural changes at \tlt\ for $x=0.135$ and
$x=0.125$ are not so dramatically different, this tells us that the anomaly in
\chic\ must be sensitive to the CO. Finally, for $x=0.155$ with its extremely weak
CO, there is no anomaly at either $T_{\rm LT}$ or $T_{\rm CO}$.

\begin{figure}[t]
\center{\includegraphics[width=0.80\columnwidth,angle=0,clip]{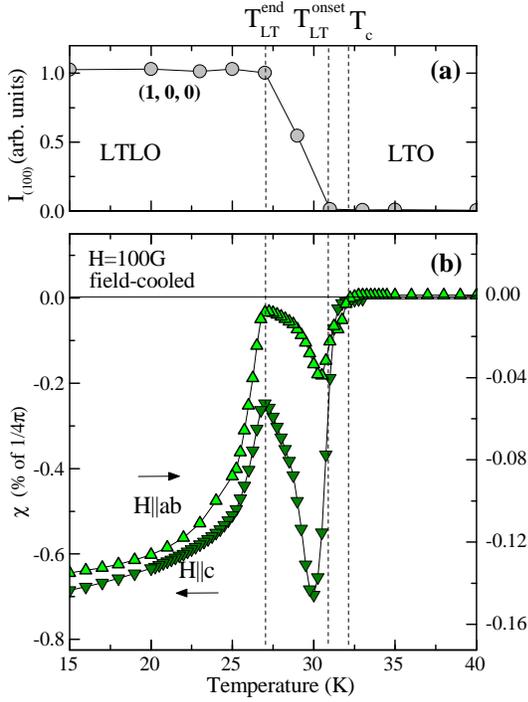}}
\caption[]{(color online) Superconductivity and stripe order in \lbcon . (a)
Integrated intensity of the (1,0,0) superstructure reflection. (b) Field cooled
signal for $H=100$~G applied parallel to the $c$-axis and the $ab$-plane. $T_c$
indicates the onset of bulk SC in the LTO phase. $T_{\rm LT}^{\rm onset}$ and
$T_{\rm LT}^{\rm end}$ denote the onset and the completion of the
LTO$\rightarrow$LTLO transition, respectively.}\label{fig11}
\end{figure}

\subsection{Superconductivity}
\label{SC}

The bulk SC phase was analyzed by magnetic \sus\ measurements. In Fig.~\ref{fig10}
we show a selection of normalized field cooled (FC) and zero field cooled (ZFC)
measurements for a magnetic field of $H=100$~G (0.01~T) applied parallel to the
$c$-axis.~\cite{pinning} Similar data sets for 2~G and 20~G reveal no additional
information. The left panels in Fig.~\ref{fig10} show how the bulk $T_c$ decreases for $x\geq
0.095$, reaching a minimum at 1/8-doping, while the right panels show how $T_c$
increases again for $x>1/8$. The bulk SC transition temperatures shown in
Fig.~\ref{fig1} were each determined from the intercept of the tangent to the steepest part
of the FC curve with $\chi=0$, for all except $x=0.135$. The latter crystal has a very
broad transition, as one can see best in Fig.~\ref{fig10}(d), which may originate
from a very steep phase boundary in that region, {\it i.e.}, large $dT_c/dx$, or sample
inhomogeneity. In addition, the crystal has a very small Meissner signal, so that
the normalization overemphasizes its diamagnetic signal with respect to the other FC
curves in Fig.~\ref{fig10}(b). Therefore, we decided to identify the $T_c$ for
$x=0.135$ with the onset temperature in the ZFC curve.

\begin{figure}[t]
\center{\includegraphics[width=0.70\columnwidth,angle=0,clip]{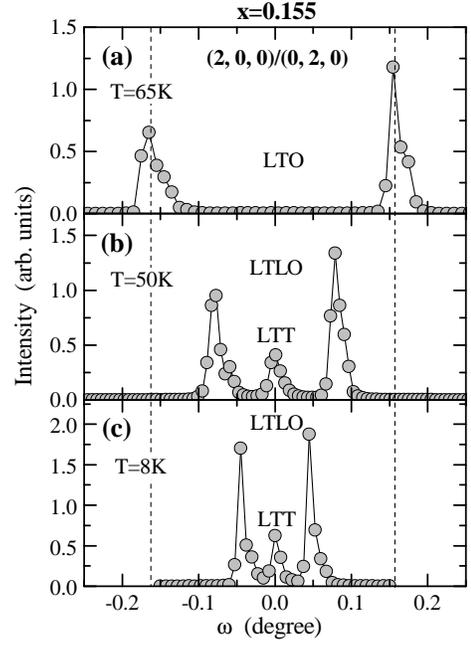}}
\caption[]{Low temperature phase transition in \lbcof . $\omega$-scans through the
(2,0,0)/(0,2,0) Bragg reflections which are simultaneously present due to twin
domains; see Fig.~\ref{fig2}(e). (a) In the LTO phase just above the phase
transition, (b) in the mixed LTLO and LTT phase  below the transition, and (c) at
base temperature. Error bars are within symbol size.} \label{fig12}
\end{figure}

\subsubsection{Special cases x=0.095 and x=0.155}
\label{special}

To properly judge the bulk SC properties of the crystals with $x=0.095$ and $x=0.155$,
we emphasize some unique features not observed for the other samples. The crystal with
$x=0.095$ is interesting, because it is the only crystal where the
LTO$\leftrightarrow$LTLO transition occurs just below the SC transition. As one can
see in Fig.~\ref{fig11}, after the initial onset of bulk SC in the LTO phase at
$T_c=32$~K, SC collapses below 30~K when both the LTO$\rightarrow$LTLO and the CO
transition occur. Once the transformation is complete, bulk SC reappears.

The crystal with $x=0.155$ is special because it is the only one which does not show
a clean structural transition to single phase LTT or LTLO. Instead, the LTO phase
transforms into a phase mixture of LTT and LTLO with a volume ratio of 1:4; see
Fig.~\ref{fig12}. Between $T_{\rm LT}$ and base temperature, the orthorhombic strain
of the LTLO phase continues to decrease monotonically; see also Fig.~\ref{fig3}. It
remains unclear whether CO exists only in the LTT phase or also in the LTLO phase.
In a study on \lbsco , static CO was observed in crystals with LTLO phase with
significantly larger remanent strain~\cite{Fujita02a}; however, the situation could
be different for $x \neq 0.125$. For these reasons the contributions of the LTT and
LTLO phase fractions to both bulk SC below $T_c$, and CO below \tco\ remain
unquantifiable for the $x=0.155$ crystal.

\section{Discussion and Conclusions}
\label{discussion}

The successful growth of \lbco\ single crystals with Ba concentrations as high as
$x=0.155$ has given us the opportunity to study the stripe phase beyond the magic
1/8-anomaly and even up to optimal doping, a region which has so far only been
accessible with polycrystalline
materials.~\cite{Moodenbaugh88a,Axe89,Maeno91,Arai03a,Vojta09a} The detection of CO
in bulk SC crystals with $x$ far below and far above $x=1/8$ is certainly the most
significant finding. The full picture, however, becomes clear only when considering
the relationship between the various properties and transition temperatures.

\begin{figure}[t]
\center{\includegraphics[width=0.94\columnwidth,angle=0,clip]{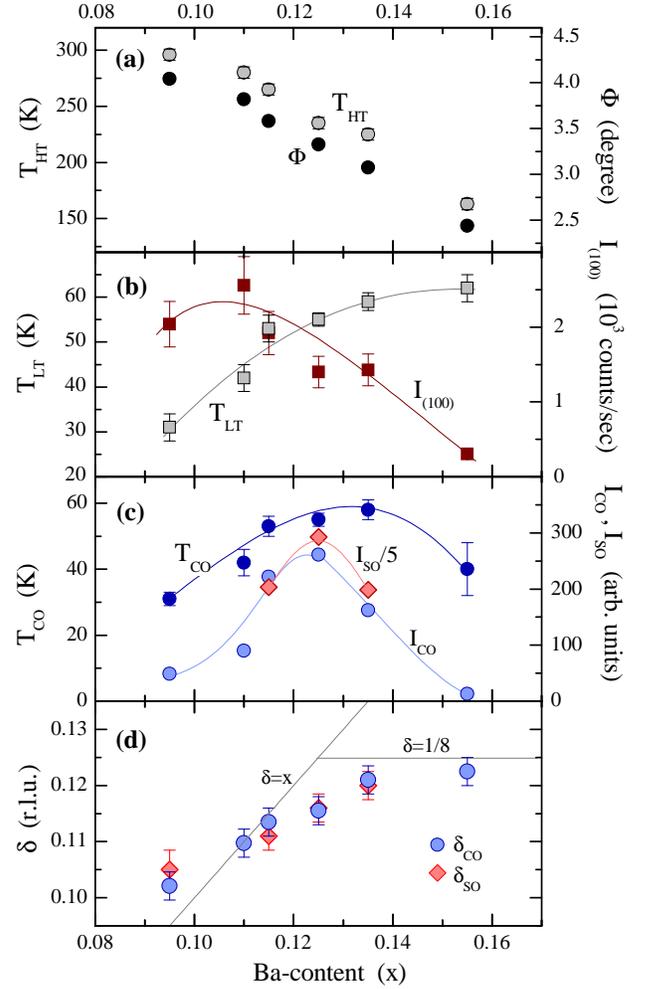}}
\caption[]{(color online) Crucial parameters of stripe phase and crystal structure
in \lbco\ versus nominal Ba content. (a) HTT$\leftrightarrow$LTO transition
temperature \tht\ and octahedral tilt angle $\Phi$ at 60~K of average structure.
$\Phi$ was calculated using $\Phi^2 = f \cdot (b_o-a_o)$ with
$f=380$~$(^\circ)^2/{\rm \AA}$ for the tilt of the apical
oxygen.~\cite{Buechner94c,Crawford05a} (b) LTO$\leftrightarrow$LTT/LTLO transition
temperature \tlt\ and integrated intensity of the (1,0,0) peak at base temperature.
(c) Charge stripe order temperature \tco\ as well as peak intensity of CO and SO
peaks. (d) Incommensurability $\delta$ extracted with XRD from the CO-peak and with
ND from the SO-peak. The solid lines $\delta=x$ and $\delta=1/8$ describe the low
and high $x$ reference of the stripe model. Solid lines in (b) and (c) are guides to
the eye. Error bars are not shown if within symbol size.} \label{fig13}
\end{figure}

\subsubsection{Variation of parameters with nominal Ba content}

A summary of important parameters versus Ba doping is given in Fig.~\ref{fig13}. In
panel (a) we compare \tht\ with the octahedral tilt angle $\Phi$ of the average
structure calculated from the orthorhombic strain just above \tlt . Besides the
monotonic variation of \tht\ and $\Phi$, one can see that stripe order occurs for
tilt angles ranging at least from 4.0$^\circ$ to 2.4$^\circ$, with stripes being
most stable at $x=1/8$ where $\Phi = 3.3^\circ$. For \lnscoxy\ a critical tilt angle
of 3.6$^\circ$ has been identified to mark a phase boundary between SC and non-SC in
the LTT phase.~\cite{Buechner94c} This boundary is not very sharp and there are no
reports yet on how deep charge stripe order persists into the SC LTT region with
$\Phi < 3.6^\circ$. The existence of such a critical angle is reasonable, since the
symmetry breaking potential of the LTT phase should scale with $\Phi$. However, in
our recent high pressure experiments on \lbco\ at $x=1/8$ we have found that static
charge stripes form even for $\Phi=0$, where the average structure has flat \pla\
planes.~\cite{Huecker10a} We believe that in this latter case the interactions
between dynamic short range charge stripe correlations and local octahedral tilts
trigger a spontaneous symmetry breaking by stripes. This mechanism may be
particularly strong for commensurate $x=1/8$ doping. It is possible that the
strength of the coupling to local displacements also depends on the local
variance~\cite{Attfield98a,McAllister02a} of the ionic radii at the lanthanum site;
that is, the critical $\Phi$ of the average LTT structure may be smaller for
compounds with a larger variance. In fact, \lbcoe\ has a larger variance than
\lnscoe , which may explain our observation of CO for $x=0.155$ with only $\Phi =
2.4^\circ$.

Further signatures of the influence of local properties of Ba are evident from
Fig.~\ref{fig13}(b), where we focus on the LTO$\leftrightarrow$LTT/LTLO transition.
In the LTT phase, the intensity of the (1,0,0) peak scales with $\Phi^2$. $\Phi$
decreases with increasing Ba content, and becomes zero in the HTT
phase.~\cite{Yamada92} In the LTLO phase, the (1,0,0) intensity also decreases with
increasing orthorhombic strain. For $x \leq 0.135$, strain at base temperature is
either zero or negligible. Hence, the observed decrease of the (1,0,0) peak towards
high doping can be naturally explained in terms of $\Phi$ for the average structure.
On the other hand, we see that \tlt\ increases with Ba doping in spite of the
decrease of $\Phi$ and \tht , thus requiring a different explanation. Here, local
distortions around an increasing number of Ba defects must be the driving force for
the transition, as has been discussed in
Ref.~\onlinecite{Attfield98a,McAllister02a}. Towards low doping, the LTT (or LTLO)
phase and, thus, the (1,0,0) peak eventually have to disappear, since there are just
not enough Ba defects to stabilize these phases. The relatively low \tlt\ and
(1,0,0) intensity for $x=0.095$ are evidence of this.

In Fig.~\ref{fig13}(c) we compare \tco\ with the peak intensities of the CO-peak and
the SO-peak. (For \tso, see Fig~\ref{fig9}). The similarity in trends for the CO and
SO phases is apparent. Both peak intensities show a maximum at $x=1/8$ and drop off
quickly for $x\neq 1/8$. In contrast, \tco\ and \tso\ describe broad domes, which do
not necessarily peak symmetrically at $x\sim 1/8$. For example, \tco\ coincides with
\tlt\ and increases up to at least $x=0.135$. In the case of \tso, our data indeed
suggest a weak maximum at $x=1/8$. This is consistent with polycrystal data from
$\mu$SR in Ref.~\onlinecite{Arai03a}, which show a clear peak at $x=1/8$, although
the \tso\ values are about 10~K lower than in our single crystals. With SO
transition temperatures as high as 42~K, the crystals' \tso\ are reminiscent of the
highest $T_c$ of \lsco\ reached under pressure when the \pla\ planes become
flat.~\cite{Yamada92}

\begin{figure}[t]
\center{\includegraphics[width=0.95\columnwidth,angle=0,clip]{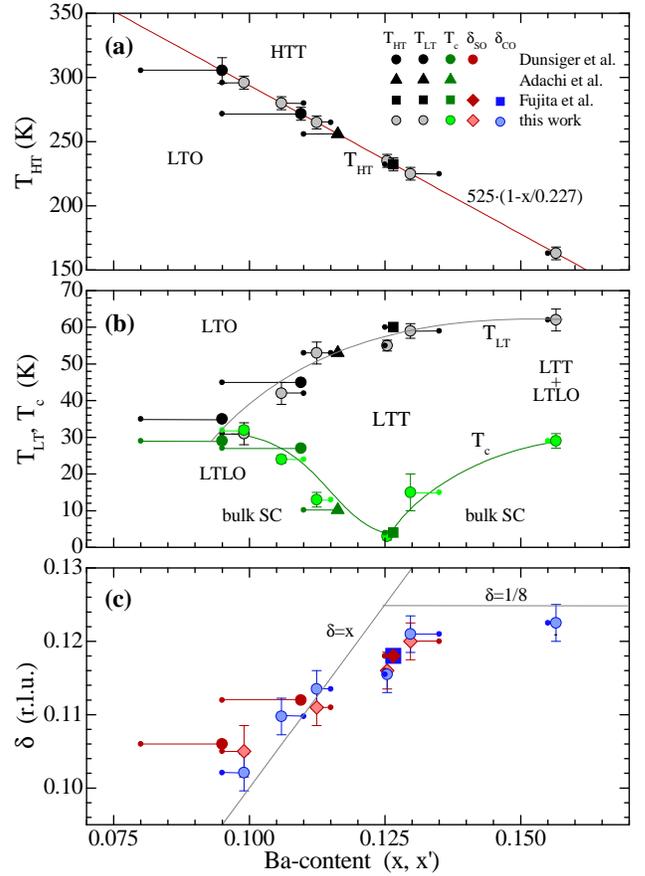}}
\caption[]{(color online) Comparison of selected parameters of \lbco\ from current
work and literature plotted versus nominal ($x$) and calculated ($x'$) Ba content.
(a) Experimental data for $T_{\rm HT}$ and theoretical curve used to estimate actual
Ba content. Small dots in (a-c) represent data plotted versus nominal Ba content,
large symbols those versus calculated Ba content. (b) Structural and bulk SC
transition temperatures \tlt\ and $T_c$. Solid lines are guides to the eye. (c)
Incommensurability $\delta$ extracted with XRD from the CO-peak and with ND from the
SO-peak. The solid lines describe the low and high $x$ reference of the stripe
model. (a-c) Where available error bars are shown. The literature data were taken
from Dunsiger et al. (Ref.~\onlinecite{Dunsiger08a}), Adachi et al.
(Ref.~\onlinecite{Adachi01a}), and Fujita et al. (Ref.~\onlinecite{Fujita04a}).}
\label{fig16}
\end{figure}

Finally, we present the incommensurability $\delta$ in Fig.~\ref{fig13}(d), for
which we find good agreement between the values determined with XRD from the CO-peak
and with ND from the SO-peak. In the phenomenological stripe model, one expects that
$\delta$ follows the solid line $\delta = x$ for $x \leq 0.125$ and saturates, or
increases less steeply, for $x>0.125$.~\cite{Tranquada97a,Yamada98a,Fujita02a} Our
data for $\delta$ match that line at $x\sim 0.11$, stay below for higher $x$, and
seem to stay above towards lower $x$. Similar deviations have been reported in
Ref.~\onlinecite{Dunsiger08a} for low $x$ and in Ref.~\onlinecite{Kim09a} for
$x=0.125$.

\subsubsection{Estimated actual Ba content; comparison with literature}

In this section, we will make the case that small discrepancies among different
studies can be reconciled to a large extent by accounting for deviations of the
actual Ba content $x'$ from the nominal $x$ value. The structural transition
temperature \tht\ is sensitive to the Ba concentration.  To use \tht\ for
calibrating $x'$, we assume that it follows the linear dependence $T_{\rm
HT}(x)=T_{\rm HT}(0) \times (1-x/x_c)$ as shown in Fig.~\ref{fig16}(a) [same as in
Fig.~\ref{fig3}(b)]. At $x=0$ this curve assumes the experimental value $T_{\rm HT}
= 525$~K for \lco\ and it goes through 235~K at $x=0.125$, which is the most
accurately known value for Ba-doped compounds. In Fig.~\ref{fig16} we compare our
data with data from literature for both nominal $x$ (small dots) and calculated $x'$
(large symbols). In Fig.~\ref{fig16}(b) and (c) one can recognize a significantly
improved agreement between the various data sets for \tlt , $T_c$, and $\delta$ when
plotted versus $x'$. In particular, for $\delta (x')$ we find a much better
agreement with the trend $\delta=x'$ for $x' < 1/8$ as shown in Fig.~\ref{fig16}(c).
Nevertheless, $\delta(x')$ still falls below  $\delta = 1/8$ for $x' \geq 1/8$,
which remains to be understood.

\subsubsection{Superconductivity and stripe order}

A key question concerns the relationship between stripe order and SC. Are stripe
correlations an essential and universal ingredient of SC in the cuprates, or just an
interesting but not crucial feature? This multifaceted problem has attracted a lot
of attention. Angle resolved photoemission spectroscopy studies show that in the
stripe ordered state \lbco\ develops a gapped Fermi surface similar to that in bulk
SC \lsco , with the antinodal gap energy $\Delta(x)$ of both groups of samples
describing a dome with a maximum at $x\sim1/8$.~\cite{Valla06a,He09a} This motivated
the idea that static stripe order does not suppress SC pairing correlations {\it per
se}, but prevents phase coherence.~\cite{Li07c} In our recent work on \lbcoe\ we
obtained further evidence for this picture. It seems that stripe order causes an
electronic decoupling of the \pla\ planes and destroys the 3D SC phase coherence,
while some kind of 2D SC fluctuations survive.~\cite{Li07c,Tranquada08a} Similar
conclusions have been reached in recent theoretical work in which the specific
stacking arrangement of stripes in \lbco\ was considered.~\cite{Berg07a,Berg09a}

If the CO and SO happened to compete with the amplitude of the SC order, then we
might expect to see a decrease in CO and SO peak intensities at the onset of bulk
SC. First we focus on the CO data in Fig.~\ref{fig5}, where we have indicated $T_c$
by vertical dash-dotted lines. The best cases to examine are $x=0.11$, 0.115, and
0.135, where $T_c$ is well below \tco\ but not too far below.  As one can see, there
is no significant change of the CO-peak intensity at $T_c$. Note that the crystal
with $x=0.155$ is not well suited for this test because of the low statistics of the
CO data as well as the presence of the LTT/LTLO phase mixture.

In the case of SO, the best candidates are the crystals with $x=0.115$ and 0.135. As
can be seen in Fig.~\ref{fig7}(b), no significant changes of SO at $T_c$ are
apparent.  [This is in contrast to Ba(Fe$_{1-x}$Co$_x$)$_2$As$_2$, where the
spin-density-wave order decreases with the onset of
superconductivity.\cite{Fernandes10}] We mention that published work by other groups
for $x<1/8$ also bears no evidence for changes of SO or CO across
$T_c$.~\cite{Dunsiger08a,Kim09a} Another question is whether the onset of SO has any
effect on the CO. It is thinkable that the onset of SO enhances the CO. However, in
Ref.~\onlinecite{Tranquada08a} we could show for $x=1/8$ that neither the intensity
nor the width of the CO-peak are affected by the SO transition. The two other
dopings where the current data allow conclusions are $x=0.115$ and 0.135 with
$T_{\rm SO}\simeq 41$~K, but also here no change of the CO at \tso\ is observed; see
Fig.~\ref{fig5}.

Overall, we find no evidence that CO and SO are affected by the onset of bulk 3D SC,
nor seems CO to be affected by the simultaneous onset of SO and weak in-plane 2D SC
correlations. Thus the coexistence of CO and SC pairing is not altered by the
development of 2D and 3D SC coherence. It seems that the defining moment for the
ultimate ground state is the CO transition itself, where depending on the hole
concentration and the discussed average and local structure parameters, the balance
between the order parameters of CO, SO, and bulk SC is determined.

\subsubsection{Comparison with LBCO, Nd-LSCO, and Eu-LSCO}

With few exceptions, our results agree well with published work on \lbco\ single
crystals and polycrystals, and have significantly expanded our knowledge on charge
and spin stripe order. As for the various critical temperatures, the largest
differences are observed between data from single crystals and polycrystals. For
example, polycrystals have significantly lower values of $T_c$ and \tso\ for a given
$x$.~\cite{Arai03a,Yamada92,Nachumi98b,Vojta09a}  Early reports on polycrystals also
show somewhat lower \tht\ and higher \tlt\ values.~\cite{Axe89} Among the available
single crystal data sets, we find good agreement when plotted versus the estimated
actual Ba content. One exception concerns the relationship between \tco\ and \tlt .
In a recent XRD study on a $x=1/8$ crystal, CO sets in significantly below $T_{\rm
LT}$, and shows a melting of the stripe stacking order before the in-plane order
disappears.~\cite{Kim08a} Here we find that $T_{\rm CO}=T_{\rm LT}$ for Ba doping up
to $x=0.135$. Only for $x=0.155$ do our XRD and magnetization data indicate $T_{\rm
CO}<T_{\rm LT}$. An early melting of the stacking order was not observed for $0.115
\leq x \leq 0.135$.

Another difference concerns the extent of the SO phase. In
Ref.~\onlinecite{Arai03a}, magnetic order, together with bulk SC, was detected by
$\mu$SR in a polycrystal with $x=0.15$; however, we find no evidence for SO in our
$x=0.155$ crystal. As long as $\mu$SR detects static order, ND should as well,
independent of a concomitant opening of a spin gap.~\cite{Kofu09a} However, the weak
CO-peak for $x=0.155$ already suggests that any SO-peak will be extremely difficult
to identify.

A comparison of Fig.~\ref{fig1} with the phase diagram of \lnsco\ in
Ref.~\onlinecite{Ichikawa00a} shows striking similarities but also important
differences. There is obviously a qualitatively similar arrangement of structural
and electronic phases, with maximum CO and SO temperatures at around $x\simeq 1/8$.
The similarity continues down to such details as \tco\ dropping below \tlt\ only for
$x>1/8$, the low-temperature structure changing to LTLO at low $x$, and a tendency
towards mixed structural phases at higher $x$, where $\Phi$ and, thus, the energetic
differences between the various possible symmetries become small. Note that for
identical $x$, $\Phi$ is smaller in \lbco , which may explain why the mixed LTT/LTLO
phase for $x=0.155$ survives down to base temperature.

A significant difference concerns the relationship between \tso\ and $T_c$. The SO
transition temperatures for the \lnsco\ single crystals in
Ref.~\onlinecite{Ichikawa00a} determined by ND are several times higher than the
maximum bulk $T_c$, with the caveat that truly static SO occurs at much lower
temperatures, as confirmed by a number of $\mu$SR
studies.~\cite{Wagener97a,Nachumi98b,Tranquada99a,Klauss00a,Savici05a} The
relatively low $T_c$ values, on the other hand, follow from a stronger suppression
in the Nd-doped system. This corresponds with a broader range of $x$ over which
stripe order is detectable. Furthermore, Nd-doping causes $T_c$ to go down even in
the LTO phase, most likely as a consequence of the larger
$\Phi$.~\cite{Buechner94c,Wagener97a,Schafer94} For comparison, in \lbco\ \tht\ and
the corresponding $\Phi$ at low $T$ are even smaller than in Nd-free \lsco .

Currently, \lesco\ with $T_{\rm LT}\sim 120$~K is the only system where $T_{\rm
CO}<T_{\rm LT}$ for $x\sim1/8$. In a recent resonant soft x-ray scattering study,
$T_{\rm CO}=80$~K and 65~K have been reported for $x=0.125$ and 0.15,
respectively.~\cite{Fink09a} The fact that these \tco\ values are significantly
higher than in \lbco\ implies that they do not solely depend on the hole
concentration, but on $\Phi$ and the local structure as well.~\cite{Simovic03a} It
also suggests that in all \lbco\ crystals with $T_{\rm CO} = T_{\rm LT}$, CO would
likely persist to higher temperatures if it were
not limited by the LTO$\leftrightarrow$LTT/LTLO transition.\\

\section{Summary}
\label{summary}
Experimental evidence for the existence of static stripe order in \lbco\ single
crystals with $0.095 \leq x \leq 0.155$ has been presented. Both the magnetic and
the charge order parameters are maximum at $x=1/8$, where bulk superconductivity is
most strongly suppressed. The competition likely involves the phase coherence of the
SC state rather than the local pairing amplitude. Neither charge order nor spin
order have shown any noticeable decrease upon the onset of bulk superconductivity.
Furthermore, charge stripe order always appears at a higher temperature than the
spin stripe order, and the charge order does not change its behavior at the onset of
spin order. Thus, charge order appears to be the leading order that both competes
and coexists with the bulk superconductivity.

\section*{Acknowledgments}

Work at Brookhaven is supported by the Office of Basic Energy Sciences, Division of
Materials Science and Engineering, U.S. Department of Energy, under Contract No.\
DE-AC02-98CH10886.  JSW and ZJX are supported by the Center for Emergent
Superconductivity, an Energy Frontier Research Center funded by the US DOE, Office
of Basic Energy Sciences. SPINS at NCNR is supported by the National Science
Foundation under Agreement No. DMR-0454672. M.H. thanks B. B\"uchner for the warm
hospitality at the IFW-Dresden where parts of the manuscript were written.


\begin{thebibliography}{80}
\bibitem[{dis()}]{displex}
\bibinfo{note}{We mention that over time we have used three displex cryostats
  and a magnet cryostat for the XRD experiments, with approximate base
  temperatures of 10~K, 7~K, 4~K, and 2.5~K, respectively. Furthermore, two
  displex cryostats have been used for the ND experiments. Although the
  temperature calibration of these instruments has been fair, we emphasize that
  the most reliable doping dependence of \tlt\ has been obtained from SQUID
  measurements.}
\bibitem[{wea()}]{weakferro}
\bibinfo{note}{The microscopic origin of the weak in-plane ferromagnetism is
  still unclear, but we mention that a similar observation has been made in
  stripe ordered \lsno\ in Ref.~\onlinecite{Klingeler05aN}.}
\bibitem[{pin()}]{pinning}
\bibinfo{note}{In 214-cuprates a strong pinning of magnetic flux is very
  common, and leads to small FC and large ZFC signals, misrepresenting the true
  SC volume, in particular in large polycrystalline samples and single
  crystals. In single crystals flux pinning depends also on microstructural
  defects and, therefore, varies with the crystal quality. For a quantitative
  comparison samples should also have identical shape and size. Since it was
  beyond the scope of the present work to achieve this for our crystals, we
  decided to normalize the FC and ZFC data.}
\end{thebibliography}

\end{document}